\documentclass[twocolumn,showpacs,amsmath,amssymb,sort,aps,pre]{revtex4-1}

\usepackage{dcolumn}
\usepackage{bm}

\usepackage{amsmath}
\usepackage{amsfonts}
\usepackage{amssymb}
\usepackage[dvips]{graphicx}
\usepackage{color}
\usepackage{ulem}
\usepackage{xcolor}

\newcommand{\Ci}{\mathrm{i}} 	
\newcommand{\e}{\mathrm{e}} 	
\newcommand{\diff}[1]{\mathrm{d}#1\hspace{0.1cm}}

\newcommand{\bra}[1]{\langle #1 \vert}
\newcommand{\ket}[1]{\vert #1 \rangle}

\renewcommand{\emph}[1]{\textit{#1}}
\newcommand{\abs}[1]{\left\vert #1 \right\vert}

\newcommand{\setR}{\mathbb{R}}

\renewcommand{\Re}{{\mathrm{Re}}}
\renewcommand{\Im}{{\mathrm{Im}}}

\newcommand{\R}[1]{\mathrm{#1}}

\newcommand{\op}{\hat}

\newcommand{\opH}{\op H}

\newcommand{\opcd}{\op c^\dagger}
\newcommand{\opcp}{\op c^{\phantom{\dagger}}}

\newcommand{\Vg}{{V_\R{g}}}

\newcommand{\Greens}{{\mathcal G}}

\definecolor{highlighted}{rgb}{0.816,0.816,1.0}

\newlength{\graphiclength}
\begin{document}





\author{A. Braun}
\affiliation{Institut f\"ur Theorie der Kondensierten Materie, Karlsruhe Institute of Technology, 76021 Karlsruhe, Germany}
\author{P. Schmitteckert}
\affiliation{Institute for Nanotechnology, Karlsruhe Institute of Technology, 76021 Karlsruhe, Germany}
\affiliation{DFG Center for Functional Nanostructures, Karlsruhe Institute of Technology, 76128 Karlsruhe, Germany}

\begin{abstract}
  We present a method to determine the impurity Greens function of the
  interacting resonant level model (IRLM) using numerical simulation techniques 
  based on the expansion of a resolvent expression in terms of Chebyshev 
  polynomials. The iterative determination of the contributions to the 
  expansion is 
  based on a Density Matrix Renormalisation Group algorithm. 
  The spectral function has lorentzian shape, where we can show that the 
  width grows monotonically with the interaction on the contact link. 
  Moreover, for values of the interaction compareable to the band width, 
  there are additional peaks showing up for energies located outside
  the conduction band. 
\end{abstract}

\title{Numerical evaluation of Greens functions 
based on the Chebyshev expansion}

\maketitle

  \section{Introduction} Despite its simplicity, the interacting resonant level model
  (IRLM), eqs.~(\ref{eq:HLead},\ref{eq:HDot}), shows some interesting, if not surprising, features. For 
  finite bias transport, a regime of negative differential 
  conductance has been found for finite electron-electron 
  interaction $U$ on the contact link 
  \cite{BoulatSaleurSchmitteckert2008,Carr_Bagrets_PS:PRL2011}. Furthermore, the 
  system shows non-monotonic behaviour  
  \cite{MehtaAndreiPRL2006,BohrSchmitteckertPRB2007,Schiller_Andrei:X2010,BranschaedelPRL2010,PhysRevB.75.125107}
  with respect to the interaction. 
  For example, the linear  conductance as a function of the gate voltage $\Vg$ has 
  Lorentzian shape, the width of which grows up to a certain 
  value of $U$. By further increasing the interaction 
  the width of the Lorentzian shrinks again \cite{BohrSchmitteckertPRB2007}.
  Besides the non interacting case, $U=0$, and the self dual point $U=2$ \cite{BoulatSaleurSchmitteckert2008,BranschaedelPRL2010,Carr_Bagrets_PS:PRL2011}, no analytic
  solution for the non-equilibrium properties are currently known, but subject to recent
   research, see \cite{FretonBoulat:PRL2014}.
  
  In the present work we now pose the question how these
  effects are reflected in the zero temperature equilibrium spectral 
  function of the interacting level. We will recover the Lorentzian shape
  for the spectrum, the width of which grows with increasing 
  interaction. However, in contrast to the linear 
  conductance, the width does not decrease again. Instead, there 
  are peaks showing up for energies \emph{outside} the 
  conduction band of the non-interacting leads.

  As a precondition to obtain the spectral function, we need to 
  compute the single particle Greens function
  of the resonant level in frequency space. 
  To this end we developed an approach based on the  expansion of the resolvent 
  in Chebyshev Polynomials of the Hamiltonian.

  In order to reach this goal we apply a numerical method based 
  on the Density Matrix Renormalization group 
  (DMRG) method \cite{White92,White93,White1996,NoackManmana2005,%
  SchollwoeckRMP2005,HallbergAIP2006}, where
  we make use of the Chebyshev expansion \cite{AbramowitzStegun,TalEzerl_Kosloff:JCP84,AlvermannFehske2006} of the function
  \begin{equation}
    f^\pm_z(x)
      =-\Ci\int_0^{\pm\infty} \diff t\,\e^{\Ci(\pm z-x)t} 
          = \frac 1 {\pm z-x},
\label{eq:function}
  \end{equation}
  with $x, \Re z\in\setR$ and $\Im(z)>0$. 
  As we will show below the main advantage of our approach lies 
  in the reconstruction of the spectral function from the moments of
  the Chebyshev expansion, where we can deconvolve a numerically inserted
  broadening ensuring that the resulting spectral function is still positive.
  The  Chebyshev approach has been
  applied before to compute the spectral function directly from 
  the series expansion of the $\delta$ function, 
  where problems due to
  Gibbs oscillations have been circumvented by means of the 
  kernel polynomial method (KPM) \cite{AlvermannFehske2006,Alvermann:Fehske:PRB2008,Alvermann_Fehske:PRL2009,Holzner:PRB2011}. 
  Gibbs oscillations appear when the
  series expansion of a discontinuous function is truncated to a finite
  polynomial order. In this work we demonstrate how to get around this
  problem without using the KPM by explicitly keeping track of a broadening of the resolvent. 
  The method we apply not merely
  allows to extract the spectral function with a well controlled 
  broadening of the spectral lines, but rather gives access to the full
  Greens function of the problem.

  \section{Interacting resonant Level model} The IRLM with tight binding leads is defined as 
  \begin{flalign}
    \opH&=\opH_\text{d}+\opH_\textsc{l}+\opH_\textsc{r}, 
    ~\opH_\textsc{r/l} = -J\!\!\sum_{x=\pm 1}^{\pm\infty}  \opcd_x\opcp_{x\pm1}\!+\!\text{h.c.}, \label{eq:HLead}\\
    \opH_\text{d} &= U \!\!\!\sum_{x=\pm 1} \!\!
        (\op n_\text{d}\!\!-\!\!{\frac1 2})(\op n_x\!\!-\!\!\frac1 2) 
         \!+\! \Vg \op n_\text{d}
         - J_\textsc c\!\!\!\sum_{x=\pm 1}\!\! \opcd_{x}\op d\!+\!\text{h.c.} , \label{eq:HDot}
  \end{flalign}
  where the dot level 
  ($\op d$, $\op d^\dagger$; $\op n_d=\op d^\dagger\op d$) is given by 
  $\opH_\text{d}$, coupled 
  to the left (L) and the right (R) lead given as 
  $\opH_\text{L/R}$ via a hopping matrix element $J_\textsc c$. 
  The interaction is restricted to a density-density 
  interaction $U$ on the contact link between the particles on
  the dot and on the first lattice site of the both leads 
  ($\op n_x=\opcd_x\opcp_x$). 
  The involved operators fulfill the common
  fermionic anti-commutation relations. 
  The energy
  of a particle occupying the dot can be shifted by applying a gate voltage
  $\Vg$. In this work, we set $\Vg\equiv0$, while considering half-filled
  leads, implying resonant tunneling at the fermi edge.
  \section{Resolvent Representation}
  The calculation of Greens functions $\Greens$ in frequency domain 
  requires the evaluation of expressions of the form 
  \begin{flalign} \label{eq:Resolvent}
    \Greens_{\op A, \op B}^\pm (z) 
       = & \bra{\Psi_0}\op A 
                  [{E_0-\op H\pm z}]^{-1}
                 \op B\ket{\Psi_0}, 
  \end{flalign} 
  $z=\omega+\Ci\eta$, $\eta>0$, where $\ket{\Psi_0}$ is an eigenstate of 
  $\op H$ with 
  $\op H\ket{\Psi_0}=E_0\ket{\Psi_0}$. 
  Based on the decomposition
  in a '$+$' and a '$-$' part, the retarded and the advanced Greens function
  in frequency representation then can be recovered as
  \begin{equation}
    \Greens^\R {r/a}_{\op A,\op B}(z)
    =\Greens^{+/-}_{\op A,\op B}(z)-\Greens^{-/+}_{\op B,\op A}(z).
   \end{equation}

  In the thermodynamic limit the convergence generating factor
  $\eta$ has to be sent to the limit $0^+$. However, the DMRG is 
  restricted to the treatment of systems with finite size which leads 
  to a finite energy level discretisation. Targetting the physics of the 
  impurity coupled to a structureless bath one therefore has to choose
  $\eta$ finite in order to average over the discrete lead levels 
  before extrapolating $\eta\rightarrow 0^+$ in order to obtain the 
  thermodynamic limit. 

  For the evaluation of \eqref{eq:Resolvent} within DMRG many techniques are available.
  The most accurate albeit expensive approach is the correction vector approach of \cite{Kuhner_White_PRB1999,Ramasesha1990}.
  There one solves a linear system for each frequency of interest. This approach allows various optimization, like a variational evaluation \cite{Jeckelmann:PRB2002}
  or the application of preconditioners to improve convergence \cite{dan06,BohrSchmitteckertPRB2007}.
  This approach even allows for changing the lattice discretization for each frequency \cite{Nishimoto_Jeckelmann:JPCM:2004,Schmitteckert:JPCS2010}.
  Variable grid discretization within variational matrix product states have been introduced in \cite{Weichselbaum_PRB2009} to obtain spectral functions
  for the single impurity Anderson model.
  Using the Lanczos method for the matrix continued fractions one can get a whole frequency range in a single DMRG run \cite{PhysRevB.52.R9827}
  which can also be implemented in an adaptive scheme \cite{Dargel:PRB20111}.
  Alternatively, one can switch to evaluating \eqref{eq:Resolvent} in the time domain combined with a Fourier transformation of the result.\cite{ PhysRevB.77.134437}
  Within the KPM approach one replaces the resolvent of \eqref{eq:Resolvent} by a delta function which is then evaluated using an expansion 
  in orthogonal polynomials where one has to apply appropriate filter kernels to ensure $A(\omega) \ge 0$.
  \section{Chebyshev expansion} 
   Here we directly expand \eqref{eq:function} 
  in terms of Chebyshev polynomials of the first kind $T_n(x)$
  \cite{AbramowitzStegun},
   \begin{equation}
    f^\pm_z(x)=\sum_{n=0}^\infty \alpha^\pm_n(z) T_n(x).
   \end{equation}
\begin{figure}[t]
    \includegraphics[width=0.95\columnwidth]{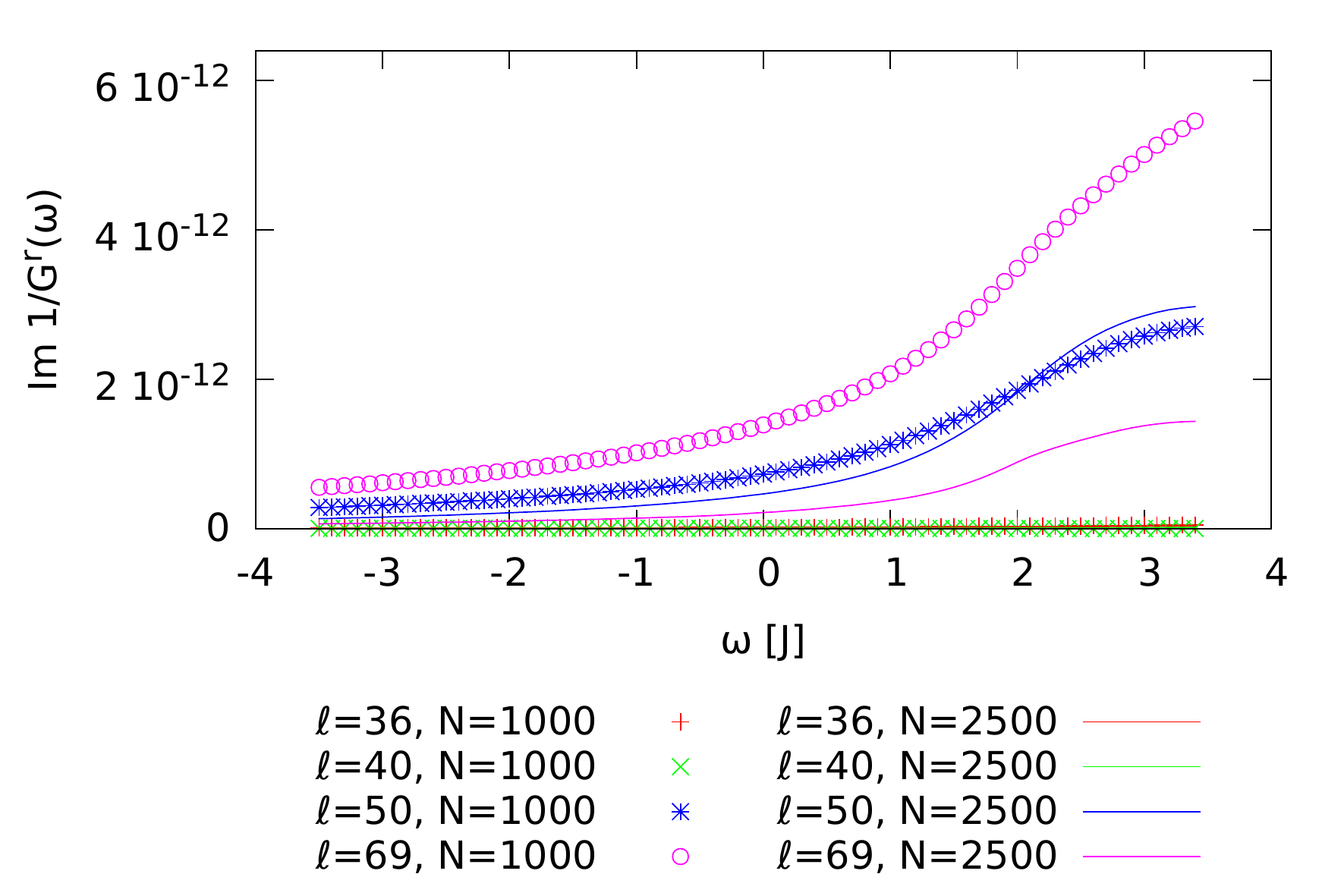}
    \caption{Imaginary part of the self energy for eigen modes $\ell = 36$, 40, 50, and 69.
       Symbols correspond to calculations using 1000 moments, lines to 2500 moments.
       In the analytic result for infinite leads the imaginary part is given by $0^+$, see eq.~\eqref{eq:GEigenMode}.
            }
\label{fig:FFimag}
  \end{figure}
The expansion coefficients $\alpha^\pm_n$ are given by
  \begin{flalign}
    \alpha&_n^\pm(z)=\frac{2/\pi}{1+\delta_{n,0}}
	  \int_{-1}^1\diff x\,
	    \frac{T_n(x)}{\sqrt{1-x^2}}\frac1{\pm z-x}
    \\
      &=\frac{-2\Ci/\pi}{1+\delta_{n,0}}
	  \int_0^{\pm\infty}\diff t \e^{\pm \Ci z t}
             \int_{-1}^1\diff x\, \frac{T_n(x)}{\sqrt{1-x^2}} \e^{-\Ci xt} 
    \\
      &=\frac{2(-\Ci)^{n+1}}{1+\delta_{n,0}}
	  \int_0^{\pm\infty}\diff t\e^{\pm \Ci z t} J_n(t) 
\label{eq:Coefficients:Bessel}
    \\
      &= \frac{2/(1+\delta_{n,0})}
	{(\pm z)^{n+1}(1+\sqrt{z^2}\sqrt{z^2-1}/z^2)^n \sqrt{1-1/z^2}},
\label{eq:Coefficients}
  \end{flalign}
  where we still keep the finite broadening $\eta$, which is essential and different to \cite{Kossloff:ARPC94}.
  We explicitly want to emphasize the relation to the Bessel functions of the first kind $J_n(t)$
  in \eqref{eq:Coefficients:Bessel}; the reason will become clear below.
  The Greens functions then can be expressed as
  \begin{flalign}
    \Greens_{\op A, \op B}^\pm (\omega) & = f_{\pm(\omega + \Ci\eta)}(\opH-E_0) \\
              &=af_{\pm a(\omega+\Ci\eta)-b}\left(a(\opH-E_0)-b\right),
  \end{flalign}
  where we chose $a,b\in\setR$ so that the spectrum of the 
  operator $a\cdot (\opH-E_0)-b$ fits into the
  interval $(-1,1)$. This rescaling has to be performed since the $T_n$ fulfill the orthogonality
  relation \cite{AbramowitzStegun}
  \begin{equation}
    \int_{-1}^1\diff x T_n(x) T_m(x)(1-x^2)^{-1/2}=\delta_{nm}(1+\delta_{n,0})/2 \,,
  \end{equation}
  therefore the spectrum has to be shifted and rescaled into the domain of $[-1, 1]$.
   Note that it is sufficient to shift the part of the spectrum, which is accessible by the the Hamiltonian,
   that is the sub space with the same quantum numbers as $\ket{\Phi_0}$ in Eq.~\eqref{eq:Phi0}.
 \begin{figure}[t]
    \includegraphics[width=0.9\columnwidth]{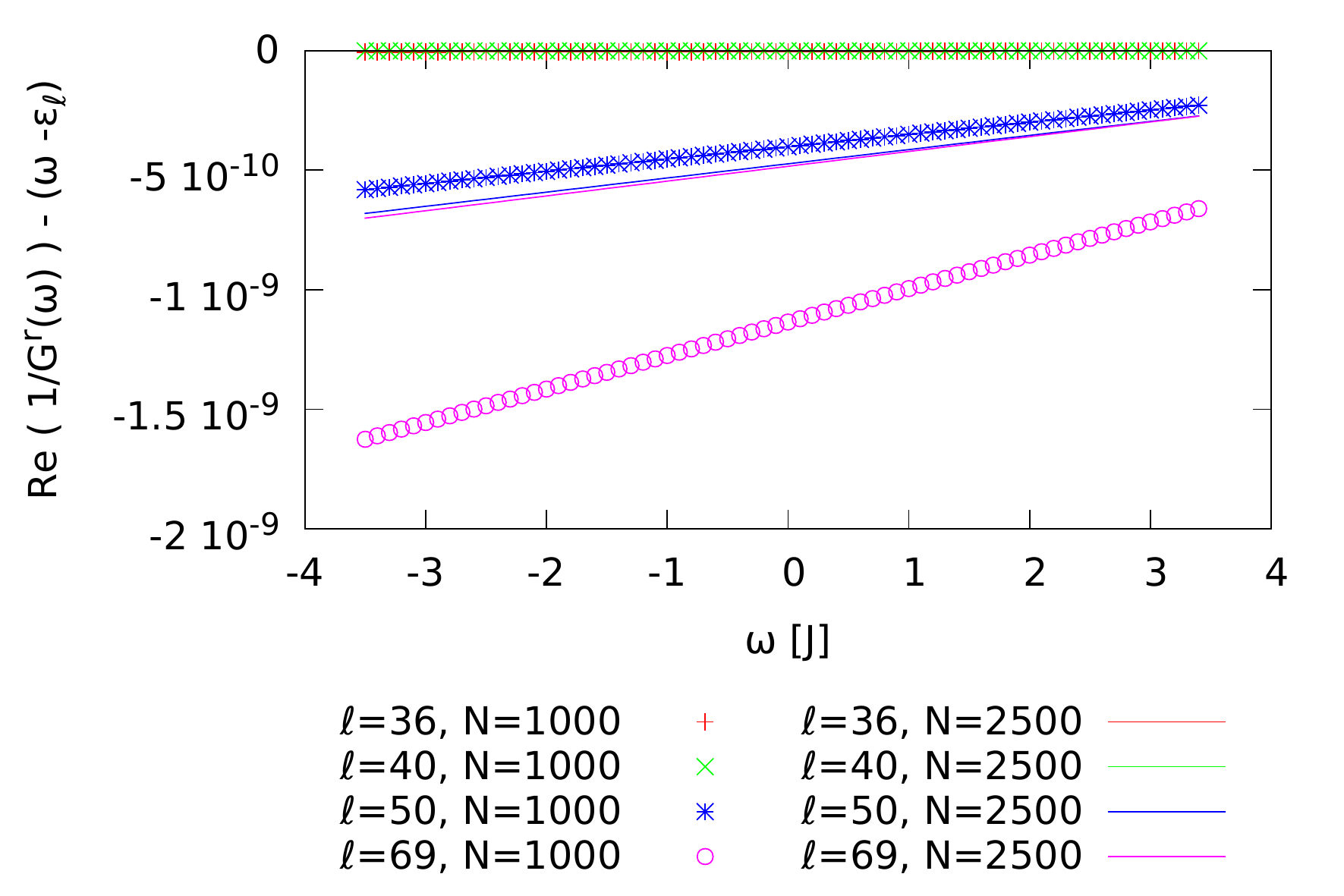}
    \caption{Deviaton of the real part of the self energy from the analytic solution \eqref{eq:GEigenMode} for eigen modes $\ell = 36$, 40, 50, and 69.
       Symbols correspond to calculations using 1000 moments, lines to 2500 moments.
            }
\label{fig:FFreal}
  \end{figure}
  Finally, for the computation of the Greens functions the Chebyshev moments $\mu_n$ 
   can be obtained numerically by evaluation of the expectation
  values of the polynomials 
   \begin{equation}
    \mu_n=\bra{\Psi_0}\op A T_n(a(\opH-E_0)-b) \op B\ket{\Psi_0},
   \end{equation}
  leading to 
  \begin{equation}
    \Greens_{\op A, \op B}^\pm(\omega) = a\sum_{n=0}^\infty \alpha^\pm_n(\pm a(\omega+\Ci\eta)-b) \mu_n.
\label{eq:expansion}
  \end{equation}
  
  To this end one exploits the recursion relation of the Chebyshev polynomials \cite{AbramowitzStegun}
   \begin{align}
      T_0(x) &= 1 \\
      T_1(x) &= x \\
      T_{n+1}(x) &=  2 x T_n(x) - T_{n-1}(x) \quad \forall \,n \ge 2 
   \end{align}

  leading to an iterative construction  \cite{AbramowitzStegun,AlvermannFehske2006} of the  $\mu_n$  using the relations 
  \begin{equation}
   \mu_n=\bra{\Psi_0}\op A\ket{\Phi_n},
  \end{equation}
 where 
   \begin{align}
      \ket{\Phi_0}     & =  \op B\ket{\Psi_0}  \label{eq:Phi0} \\
      \ket{\Phi_1}     & =  [a(\opH-E_0)-b]\ket{\Phi_0} \\
      \ket{\Phi_{n+1}} & = 2[a(\opH-E_0)-b]\ket{\Phi_n}-\ket{\Phi_{n-1}}.
   \end{align}
 
For the numerical computation of $\Greens^\pm$, the
  expansion in Eq.~\eqref{eq:expansion} has to be truncated to a finite number of contributions, 
   \begin{equation}
    \Greens^\pm \propto \sum_{n=0}^\infty \alpha^\pm_n \mu_n\approx\sum_{n=0}^N \alpha^\pm_n \mu_n.
 \label{eq:expansion_N}
   \end{equation} 
  It is now important to determine a suitable value for $N$ in order to obtain reliable numbers.
  It is known that $-\mathcal N^2 \leq \mu_n \leq \mathcal N^2~\forall n$, where 
  $\mathcal N^2=\bra{\Psi_0}\op A\op B\ket{\Psi_0}$. Therefore it is sufficient to study the behaviour
  of the coefficients $\alpha^\pm_n(z)$ for values of $z$ in the desired range. A very simple estimate can
  be given by looking at the derivation of the $\alpha^\pm_n$, and 
  especially by considering the 
  properties of the Bessel functions in Eq.~\eqref{eq:Coefficients:Bessel}: For $n>\abs t$, $J_n(t)$ rapidly 
  drops to zero with growing $n$, relating the maximum time $\abs t$ to $N$, 
  so that $N \gtrsim \abs t$. The maximum time, on the other 
  hand, is determined by $\Im(z)$, the latter exponentially cutting off the infinite integration range. 
  Therefore, we get an estimate $t \approx 1/\Im(z)$, which finally leads us to the expression
  \begin{equation}
    N \gtrsim (a\eta)^{-1}.
\label{eq:estimate_N}
  \end{equation} 
  This estimate also shows a certain limitation of the approach. Since $a$ is proportional to the inverse
  difference of the extremal eigenvalues of the Hamiltonian of the system, $1/a$ grows with the system
  size if the filling factor is kept constant, and hence the number of moments $\mu_n$ that has to be
  computed.
  On the other hand, resolving narrow structures in the spectrum requires $\eta$
  to be chosen small as compared to the width of the spectral structure of interest.

    From this follows the main disadvantage of the Chebyshev expansion, namely that the resolution is inverse proportional
  to the many-particle spectrum. Therefore, one has to increase the number of moments proportional
    to the system size in order to achieve the same resolution. In \cite{Holzner:PRB2011} the KPM 
    was extended with a  Krylov subspace type energy projector to avoid this required increase of
    moments. However, the projector itself is not cheap. E.g. in \cite{Holzner:PRB2011} a Krylov
    subspace of order 30 was applied which corresponds roughly to the cost of evaluating 30 moments.
    Since the projector is needed for each moment, we can afford for 30 times the moments for the same price tag.
    In addition, as every iterative method, the projector approach risks loosing  an interesting part of the spectrum.
    Nevertheless, the projector technique could be combined with the approach presented in this manuscript
    and the performance of the approaches may depend on the models under consideration. 
    
    Concerning the DMRG approach we would like to point out, that in contrast to \cite{Holzner:PRB2011} we are not using an 
    adaptive DMRG scheme, where one only keeps the three states needed for a Chebyshev iteration in a DMRG sweep.
    Instead we perform all Chebyshev iterations in each DMRG step and construct a density matrix out of all Chebyshev vectors
    in order to optimize the target space basis. Specifically, we first perform a DMRG run performing two Chebyshev iterations only.
    We then iteratively restart the DMRG calculation increasing the number of Chebyshev iterations performed in each run.
    Note that the use of an adaptive scheme \cite{White04,*Daley04,Dargel:PRB20111} is straightforward.
    Or one can combine both approaches, similar to the time evolution strategy in \cite{BoulatSaleurSchmitteckert2008},
    using a full DMRG strategy putting all moments in a density matrix for the first few Chebyshev iterations, and then switching
    to an adaptive scheme. 
    
    Finally we would like to address the difference to the KPM scheme. In the KPM scheme one first takes the $\eta \rightarrow 0$ limit of
    imaginary part of \eqref{eq:function} in the continuum limit, resulting into a delta function in energy. Therefore the KPM 
    provides the spectral function only, while our approach gives  the real and the imaginary part of the Greens function. 
    More importantly, our approach allows us to stay in the physically correct limit, namely first taking the system size to infinity, 
    and then $\eta$ to zero, 
    e.g.\ see \cite{Schmitteckert:JPCS2010}.
    On finite size systems this correspond to taking the level broadening $\eta$ larger then the level spacing of the system.
    Specifically, in the extrapolation scheme described in the next section, we will always employ broadening values in the physically
    correct limit.
    
\section{``Poor man's'' deconvolution}
  The zero-temperature ground-state spectral function of the 
  impurity $\mathcal A$ is related 
  to the retarded impurity Greens function by
   \begin{equation}
    \mathcal A(\omega)=-\frac 1 \pi \lim_{\eta\rightarrow 0^+} \Im \Greens^r(\omega+\Ci\eta),
   \end{equation}
  where
    $\Greens^r(z)
      =\Greens^+_{\op d,\op d^\dagger}(z)-\Greens^-_{\op d^\dagger,\op d}(z)$,
  evaluated for the ground state wave function $\ket{\Psi_0}$ of
  the system at half filling. Note that the approach presented in this work allows to solve 
  the resolvent equation \eqref{eq:Resolvent} for
  any reference state $\ket{\Psi}$. However, only for eigenstates of the Hamiltonian
   it corresponds to the Fourier transform of a  Greensfunction defined in time domain.
  
  For a system with continuous spectrum, $\mathcal A$ will be a continuous 
  function of the energy, while for a system with a discrete spectrum, 
  it will show sharp $\delta$ peaks for the discrete eigen-energies of
  the system. Now, the models we want to consider describe a nanostructure 
  coupled to \emph{infinite leads}, the latter providing for a continuous
  spectrum, while in contrast, the models we implement for the numerical
  simulation are \emph{finite}, with an overall number of $M$ lattice
  sites, leading to a discretisation of the energy spectrum. In order to
  obtain an approximation to the thermodynamic limit, we therefore choose 
  the convergence generating factor $\eta$ to be finite, which leads to 
  averaging over a few discrete energy levels. A lower bound 
  is given by $\eta\gtrsim 2 \pi J / M$, the level  spacing of the non-interacting leads.
  
  The finite value of $\eta$
  adds an artificial broadening of the energy levels.
  In \cite{Schmitteckert:JPCS2010}, a method to remove this broadening was 
  introduced. Based on the assumption that the self-energy 
  $\Sigma(\omega)=\omega+\Ci0^+-[\Greens^r(\omega)]^{-1}$ is shifted by 
  $\Ci \eta$ as compared to the result for the thermodynamic limit, 
  $\Sigma_\eta(\omega) = \omega+\Ci 0^+ -[\Greens^r(\omega+\Ci\eta)]^{-1}\stackrel!=\Sigma(\omega)-\Ci\eta$, 
  a sharpened Greens function can be computed directly. This was 
  succesfully checked for the energy eigenstates of a tight binding 
  chain of free fermions, where the relation holds exactly.
  In general, the broadened self-energy will depend on $\eta$ in a more 
  complicated way, which can spoil the approach. For example, for the 
  impurity Greens function of the RLM with tight binding leads, the 
  self-energy depends non-linearly on the broadening $\eta$ -- including 
  the real part of $\Sigma_\eta$. Therefore we generalize the assumption 
  where we now take the self-energy as a function of the broadening,
  \begin{equation}
    \Sigma_\eta(\omega)=\Sigma^{(0)}(\omega)+\Sigma^{(1)}(\omega)\eta+\Sigma^{(2)}(\omega)\eta^2+\ldots,
\label{eq:Greensfn:SelfEnergyEtaFit}
  \end{equation}
  allowing for an extrapolation to $\eta=0$ from numerical data with 
  finite broadening. 
  The Chebyshev
  expansion of the Greens function allows to obtain the self-energy for 
  many different values of $\eta$ and $\omega$ based on a single 
  calculation for the moments $\mu_n$, since both, $\eta$ and $\omega$,
  only enter via the $\alpha_n$  of Eq.~\eqref{eq:Coefficients},
  which makes this approach cheap 
  as compared to the correction vector method, where the calculation has 
  to be repeated for every single pair of values of $\eta$ and $\omega$. 
  The self-energy for the thermodynamic limit can then be identified as
    $\Sigma(\omega)\equiv \Sigma^{(0)}(\omega)$,
  which, in turn, yields the value of the Greens function in the 
  thermodynamic limit. 
  Specifically, according to  eq.~\eqref{eq:Greensfn:SelfEnergyEtaFit} 
  we perform a polynomial fit in $\eta$ independently for the real and imaginary part
  of $\Sigma_\eta(\omega)$ for every frequency point $\omega$ of interest
  using the {\tt gsl\_multifit\_linear} function of the Gnu scientific library {\tt libgsl} \cite{GSL}.
  
   For a further discussion, like the evaluation of time dependent Greens functions or the use of Laguerre polynomials,
   we refer to \cite{Braun:2011}.
   
   Finally we note that the Chebyshev moments $\mu_n$ are typically strongly oscillating with respect to the moment index $n$.
   Therefore, the final results oscillate a little bit when changing the maximal Index $N$ slightly. In return we find small oscillating
   parts in the the spectral funtcion, if we choose $N$ to small. However, this can be avoided by implementing a smoothing window,
   e.g.\ a $\cos^2$ filter,
   for the last $N_{\rm S}$ moments. In that way one can obtain a good estimation for the spectral function with a smaller number
   of moments compared to Eq.~\eqref{eq:estimate_N}.
   
 \section{Resonant level benchmark}
 We first benchmark our approach against the Greens function of an eigen mode
 for a homogeneous tight binding chain with a nearest neighbor hopping of $J=1$ consisting of $M$ sites
 with hard wall boundary conditions corresponding to an isolated lead $\opH_\textsc{r/l}$
  in \eqref{eq:HLead}. The Greens functions for the $\ell$-th eigen mode corresponding to
  \begin{equation}
    \hat f_{\ell} = \sqrt{ \frac{2}{M+1} } \, \sum_{x=1}^{M} \sin\left( \frac{ \ell \pi x}{ M+1} \right)  \,  \hat c_x
  \end{equation}
is given by
  \begin{equation}
    \Greens^\text{r}_{\ell}  = \frac{ 1}{ \omega - \varepsilon_\ell + \Ci 0^+} \, \label{eq:GEigenMode}
  \end{equation}
  with $\varepsilon_\ell = -2 J \cos\left( \ell \pi / (M+1) \right)$.

   In Fig.~\ref{fig:FFimag} we show the imaginary part of $[\Greens^r(\omega)]^{-1}$ 
   for a half filled 70 site system using $N=1000$ (2500) moments of the Chebyshev expansion. 
   A quadratic $\eta$-extrapolation was performed using  20 different values starting with $\eta=0.6$ ($\eta=0.3$) increasing each $\eta$ value by a factor of 1.3.
   A cubic extrapolation gives basically the same result.
   In Fig.~\ref{fig:FFreal} we show the deviation of the real part of the inverse Greens function from the exact solution \eqref{eq:GEigenMode}.
   Noting that the accuracy of the diagonalization within the DMRG was restricted to a tolerance of $10^{-9}$ the results show
   that within numerical accuracy we have found the exact solution corresponding to a $\delta(\omega - \varepsilon_\ell)$ spectral function of eq.~\eqref{eq:GEigenMode}
   in the complete range starting near the Fermi point ($\ell=35$), the mid of the band ($\ell=40$, $50$), and at the band edge ($\ell=69$).
  Within the DMRG the discarded entropy was below $10^{-9}$. To ensure this we had to keep at most $m=670$ states per DMRG block for $\ell=36$, 40, and 50,
  and up to $m=1950$ states for $\ell=69$. Note that the discarded entropy is given by $- \sum_{n>m}  \alpha_n \log (\alpha_n)$, where the sum runs over the 
  the eigenvalues of the reduced density matrix of the states which are not kept.
  
  At first sight the results presented in Figs.~\ref{fig:FFimag}, \ref{fig:FFreal} may appear boring.
  However, we would like to emphasize that the error in the real part of the self energy should be compared to the values of the single particle levels $\epsilon_\ell$,
  while the error in imaginary part directly gives additional broadening introduced by the method. In addition we would like to stress that the imaginary part of the self energy,
  while being very small, is positive.
  In addition we would like to point out, that in a very recent work \cite{Jeckelmann:PRB2014} the spectral function for
  a non interacting reference system showed a spurious peak and much stronger deviations in the spectral function, see Fig.~1 in \cite{Jeckelmann:PRB2014}. 
  This is also in strong contrast to the Gibbs oscillations and the pretty large deviations  that appeared in the KPM based DMRG approach, see Fig.~3a of \cite{Holzner:PRB2011}.
  There a peak height of a spectral function corresponding to a $\delta$-function is reported as $16.23$, implying a width of the order of $1/16.23 \approx 0.06$,
  since the area of the $\delta$-function is  one. This clearly demonstrates that our approach is indeed different from the KPM.
  Especially the absence of the Gibbs oscillations is an important advantage of our approach.

  \begin{figure}[t!]
    \includegraphics{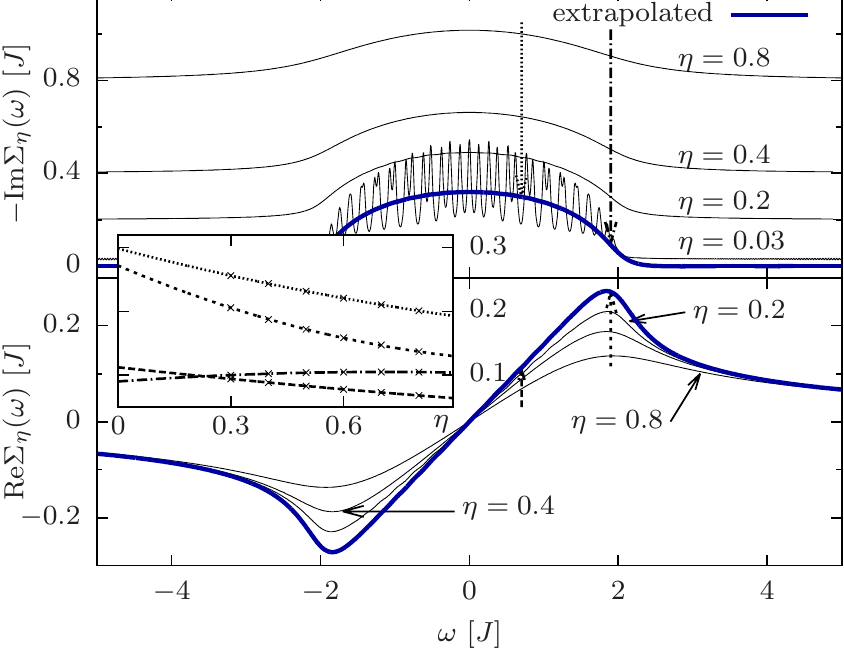}
    \caption[Broadening dependent self-energy]
            {Imaginary and real part of the broadening-dependent 
             self-energy for different values of the convergence generating 
             factor $\eta$. The thick lines represent $\Sigma^{(0)}$ as 
             defined in eq.~\eqref{eq:Greensfn:SelfEnergyEtaFit}, for a fit 
             up to quadratic order.
              The system consist of $M=48$ lattice sites in total, the 
              impurity is coupled to the leads via $J_\R C=0.4J$. 
              For $\Im\Sigma$ we include values for $\eta=0.03$ in order to 
              demonstrate the effect of the level discretisation. 
	     The inset 
             shows $\Re\Sigma_\eta$ and $-(\Im\Sigma_\eta + \eta)$ as 
             function of $\eta$ for selected values of $\omega$, compare
             the vertical arrows in the main plot.
            }
\label{fig:Greensfn:SelfEnergy}
  \end{figure}
\section{Spectral function of the IRLM}
  Having established our deconvolution scheme in combination with our DMRG approach
  we now turn to the resonant level model. In Fig.~\ref{fig:Greensfn:SelfEnergy} we show the 
  real and the imaginary part of the self-energy, for the impurity Greens 
  function $\Greens^\text{r}$ of the RLM. The numerical data have been 
  obtained from the simulation of a system with $M=48$ lattice sites in 
  total, for a single dot coupled to the leads via $J_\textsc c=0.4J$, with 
  vanishing interaction $U\equiv 0$. The different curves correspond to 
  different values of
  the convergence generating factor $\eta$. We include $\eta=0.03$ which makes the discretization
  of the single particle energy levels, due to the finite size of the system, visible. Including only
  those data points for the extrapolation procedure that do not show the finite size discretization
  then yields the result for the thermodynamic limit. 
  The inset demonstrates
  the extrapolation procedure where we explicitly show the $\eta$ dependence of $\Re\Sigma_\eta$ and 
  $\Im\Sigma_\eta$, for two different values of the frequency $\omega$. The lines correspond to the
  fit in quadratic order, where the line style relates the inset to the vertical arrows which indicate
  the corresponding value of $\omega$. Note, that the numbers are based on $N=4000$ Chebyshev moments
  $\mu_n$ obtained using a DMRG algorithm where all $\ket{\xi_n} =  T_n(a(\opH-E_0)-b) \op B\ket{\Psi_0}$ wave functions
  are added to the reduced density matrix, similar to the full td-DMRG described in \cite{Schmitteckert04}.

   For $M=48$ lattice sites at half filling, the extremal 
  eigenvalues of $\op H$ enforce a rescaling factor $a\lesssim 0.033J^{-1}$,
  where we have chosen $a\approx0.0169J^{-1}$. Using 
  eq.~\eqref{eq:estimate_N}, this leads to a minimal value for the convergence generating factor of 
  $\eta\gtrsim 0.015$. 

  \begin{figure}[t!]
    \includegraphics[width=8.5cm]{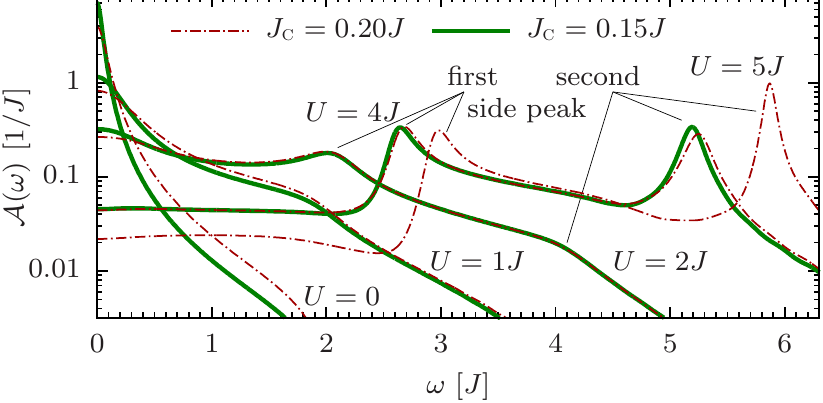}
    \caption[Spectral function of the IRLM]
            {Spectral function of the IRLM
             for different $U$. 
              The system 
              consists of $M=96$ ($M=168$) lattice sites in total, the 
              impurity is coupled to the leads via 
              $J_\R C=0.2J$ ($J_\R C=0.15J$).
            }
\label{fig:Greensfn:SpectralFunctionIRLM}
  \end{figure}

  \begin{figure}[b!]
    \includegraphics[width=8.5cm]{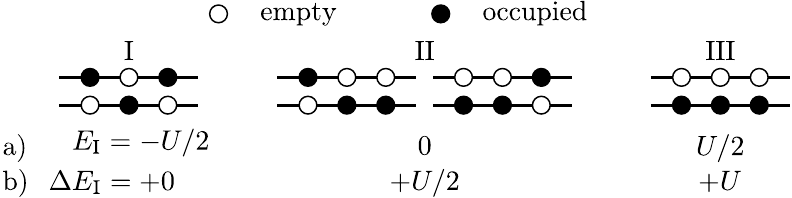}
    \caption[Basis states]
            {Basis states of the interacting region in terms of occupation 
             number eigenstates, including the dot site as well as the first 
             lead site of the left and the right lead.
             a) Interaction contribution $E_\textsc i$ to the energy of the 
                state.
             b) Interaction contribution $\Delta E_\textsc i$ to the energy 
                difference between the excitation and the state with lowest 
                energy.
            }
    \label{fig:Basis}
  \end{figure}
  Using the method defined above it is now possible to obtain $\mathcal A$ 
  for the thermodynamic limit with $\eta\rightarrow 0^+$. In 
  Fig.~\ref{fig:Greensfn:SpectralFunctionIRLM} we show results for two 
  different couplings $J_\textsc c=0.2J$ ($J_\textsc c=0.15J$), based on a simulation
  of $N=500$ ($N=800$) Chebyshev moments for a system with $M=96$ 
  ($M=168$) lattice sites in total. Due to broad peaks the number of moments are sufficient.

  Within the DMRG simulations the discarded entropies where below $4 \cdot 10^{-4}$ for the 96 and 168 site systems, considerably smaller for the small $U$ values,
   keeping up to 5600 states per DMRG block. Note that the discarded entropy is a measure of the discarded entropies including all wave functions
   constructed by the $N$ moments.
  
   In the wide band limit, i.e.\  $J_\textsc c \ll J$,  the spectral function of the noninteracting RLM is given by a Lorentzian, see eq.~\eqref{SF:RLM}.
   Increasing $U$ has two effects: first, the central peak
  of the spectral function gets broadened. For values of 
  $U\lesssim 2J$, the peak survives while for 
  $U\gtrsim4J$, it seems to disappear completely. Reducing the 
  coupling $J_\textsc c=0.2J\rightarrow0.15J$ leads to an increased 
  height and a reduced width of the central peak, which leads us 
  to the assumption that in the limit of very small $J_\textsc c$, 
  the central peak could survive for values of $U>4J$. 
  The position of the side peaks seems not to be influenced by 
  $J_\textsc c$. Their emergence can be interpreted based on the 
  sketch of the possible contributions to $\ket{\Psi_0}$ in
  Fig.~\ref{fig:Basis}. Here, we show the eight different possibilities to 
  occupy the dot as well as the neighbouring lattice sites on the leads,
  together with a) the interaction contribution $E_\textsc i$ to the energy
  as well as b) the corresponding difference $\Delta E_\textsc i$.
  For $U=0$, the ground state occupation of these lattice sites 
  is completely guided
  by the single particle behaviour as well as the Fermi statistics, leading 
  to the typical Lorentzian shape of the spectral function corresponding
  to a finite life time of an excitation on the dot level. For
  moderate values of $U \gtrsim J$, the degeneracy of the different 
  contributions to $\ket{\Psi_0}$ with respect to the 
  interaction energy $E_\textsc i$ gets lifted, leading to a splitting of 
  the resonance peak corresponding to the energy difference 
  $\Delta E_\textsc i$. The central peak loses weight because the most
  unfavourable type III contributions to $\ket{\Psi_0}$ are being suppressed. 
  For $U\gg J$, the only remaining contributions to
  $\ket{\Psi_0}$ are of the type I, while both type II and III contributions get
  suppressed. This explains the shrinking weight of the first side peak as
  compared to the second one, which also can be observed for growing $U$ on 
  Fig.~\ref{fig:Greensfn:SpectralFunctionIRLM}.

\begin{figure}[t!]
    \includegraphics[width=8.5cm]{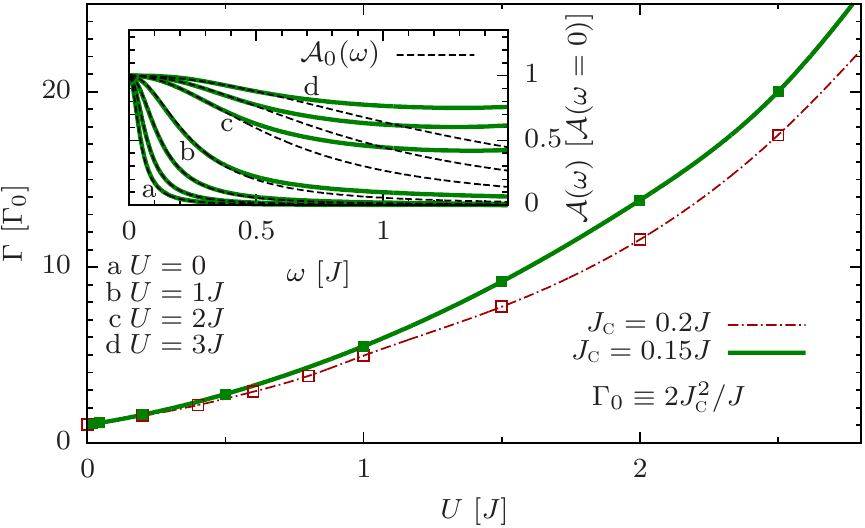}
    \caption[Width of the spectral function for the IRLM]
    { Inset: Central peak of the spectral function if the IRLM
      normalized to the value for frequency $\omega=0$.
      for different values of the interaction,
      a: $U=0$, b: $U=1J$, c: $U=2J$, d: $U=3J$.
      Unlabbeled lines correspond to interaction values in beween, see main figure.
      The system consists of
      $M=168$ lattice sites in total, the dot is coupled to the leads via
      $J_\textsc c=0.15J$. The dashed lines correspond to a fit of the 
      non-interacting wide-band limit $\mathcal A_0(\omega)$ to the 
      numerical data (solid lines). Main figure: Width $\Gamma$ of the 
      central peak as function of $U$.
       for two different values of the coupling $J_\textsc c=0.15J$, $0.2J$,
       normalized to the width in the wide-band limit $\Gamma_0 = 2J_c^2 /J$.
         For  $J_\textsc c=0.15J$ the datapoints correspond to the lines of the inset.
       Increasing $U$ leads  to monotonous growth of $\Gamma$. 
      Lines are guides to the eye.
    }
\label{fig:Greensfn:WidthIRLM}
  \end{figure}

  The broadening of the central peak is represented in 
  Fig.~\ref{fig:Greensfn:WidthIRLM}. On the inset
  we show the central peak of the spectral function 
  $\mathcal A$, normalised to the maximum value 
  $\mathcal A(\omega=0)$, for values of the interaction 
  $U=0\ldots 3J$.
  The dashed curves correspond to a Lorentzian,
   \begin{equation}
     \mathcal A_0(\omega)
 	=\frac 1\pi \frac{\Gamma}{\Gamma^2+\omega^2}.  \label{SF:RLM}
   \end{equation}
  For the RLM with $U=0$, this expression 
  corresponds to the wide-band limit of the spectral function
  $\mathcal A$, where the width 
  $\Gamma\equiv\Gamma_0=2J_\textsc c^2/J$ is determined by the 
  coupling $J_\textsc c$. We now fit this expression to the central
  peak of the numerical data for $\mathcal A(\omega)$ in 
  order to obtain its width 
  $\Gamma$. In the non-interacting case a, we find good 
  agreement of the numerical data and the wide-band limit 
  $\mathcal A_0$, indicating that for the precise value of the 
  coupling ($J_\textsc c=0.15J$), band curvature effects do not play
  a major role, at least for vanishing interaction. The same 
  still holds true for finite, increasing interaction, as long 
  as $U\ll J$. For values of $U\gtrsim J$, we find
  strong deviations from the Lorentzian shape; cf. also 
  Fig.~\ref{fig:Greensfn:SpectralFunctionIRLM}. Nevertheless,
  the width $\Gamma$ is still well defined for a small region 
  at the Fermi level. However, for values of the interaction 
  $U>3J$, as discussed before, the central peak vanishes 
  completely, rendering the width ill-defined.

  The behavior of the width $\Gamma$ depending on the 
  interaction $U$, normalized to the width 
  $\Gamma_0=2J_\textsc c^2/J$ of the noninteracting RLM, is 
  represented on the main panel. We clearly find monotonous
  growth of $\Gamma$ until the point where the central peak 
  vanishes. Interestingly, $\Gamma$ does not show any noticeable
  behavior when passing the self-dual point $U=2J$, where 
  certain non-equilibrium problems can be solved analytically
  \cite{BoulatSaleurSchmitteckert2008,BranschaedelPRL2010}, and
  where the linear conductance obtains its maximum width 
  \cite{BohrSchmitteckertPRB2007}. Unfortunately, the plot 
  for $\Gamma(U)$ can not be continued beyond 
  $U\approx 3J$ based on the available data due to the 
  vanishing central peak. 
  While we can not provide a simple physical picture on this discrepancy
  in the behaviour of the the transport properties and the spectral function
  with respect to $U$, it is plausible that the pairing, effective charge of $2e$,
  in the low voltage regime, and the charge fractionalization, effective charge $e/2$,
  in the large voltage regime, see \cite{BranschaedelPRL2010,Carr_Bagrets_PS:PRL2011}, is not reflected
  in a single particle quantity, even if it is a dynamical one.
  

    %
  \section{Summary} To conclude, we have discussed a method to obtain the 
  complete Greens function and in particular the spectral function 
  of an interacting nano structure from the Chebyshev
  expansion of the resolvent operator, based on the DMRG. We have related the 
  energy resolution to the truncation order of the series 
  expansion, which constitutes the main limitation of the approach. 
  The advantage over the KPM consists in the good control 
  over finite size effects. E.g. we could reconstruct the $\delta(\omega)$  spectral function
  of  eigen modes of a tight binding chain. Specifically, we
  can then test various extrapolations schemes (order of fitting, fitting range, smoothing of the higher order moments)
  after the expensive DMRG calculations, as none of these parameter enter the calculation of the moments.

  The method was successfully applied to the IRLM, revealing 
  interesting features of the ground state wave function. 
  In contrast to a many-lead expansion \cite{PhysRevB.75.125107} we find a monotonic dependence of
  the width of the low frequency impurity spectral function with respect to the interaction $U$.
  Finally we want
  to emphasize that the presented method is not restricted to
  the IRLM and also can be applied to any model accessible within DMRG, e.g.\ the calculation of
  the spectral function of Majorana edge states.\cite{Thomale:PRB2013}
  
 \acknowledgments
 PS would like to thank Andreas Alvermann and Holger Fehske for many discussions concerning the application of the KPM.
\bibliographystyle{apsrev}

\bibliography{bibliography}

\end{document}